\documentclass[final,5p,times,twocolumn,sort&compress]{elsarticle}

\usepackage{epsfig}
\usepackage{amsmath}
\usepackage{url}
\usepackage[colorlinks,linkcolor=blue,anchorcolor=blue,citecolor=blue,urlcolor=blue]{hyperref}
\usepackage{graphicx}
\usepackage{lineno}
\usepackage{color}
\usepackage{ulem}
\usepackage[marginal]{footmisc}

\journal{Science Bulletin}

\begin{document}

\begin{frontmatter}

\title{A new approach for deducing $rms$ proton radii from charge-changing reactions of neutron-rich nuclei and the reaction-target dependence}

\author[bh]{JiChao Zhang}

\author[bh]{Bao-Hua Sun\corref{cor1}}
\ead{bhsun@buaa.edu.cn}

\author[bh,RCNP]{Isao Tanihata\corref{cor1}}
\ead{tanihata@rcnp.osaka-u.ac.jp}

\author[SaintMary,TRIUMF]{Rituparna Kanungo}

\author[GSI,Justus-Liebig,HFHF]{Christoph Scheidenberger}

\author[bh]{Satoru Terashima}

\author[bh]{Feng Wang}

\author[GSI]{Frederic Ameil}
\author[SaintMary]{Joel Atkinson}
\author[Santiago]{Yassid Ayyad}
\author[IIT]{Soumya Bagchi}
\author[Santiago]{Dolores Cortina-Gil}
\author[GSI,Justus-Liebig]{Iris Dillmann}
\author[SaintMary,GSI]{Alfredo Estrad\'e}
\author[GSI]{Alexey Evdokimov}
\author[GSI]{Fabio Farinon}
\author[GSI,Justus-Liebig]{Hans Geissel}
\author[GSI]{Giulia Guastalla}
\author[Comenius]{Rudolf Janik\fnref{1}} 
\author[SaintMary,Dalhousie]{Satbir Kaur}
\author[GSI]{Ronja Kn\"obel}
\author[GSI]{Jan Kurcewicz}
\author[GSI]{Yury Litvinov}
\author[GSI]{Michele Marta}
\author[Santiago]{Magdalena Mostazo} 
\author[GSI]{Ivan Mukha} 
\author[GSI]{Chiara Nociforo}
\author[RCNP]{Hooi Jin Ong}
\author[GSI]{Stephane Pietri}
\author[GSI]{Andrej Prochazka}
\author[Comenius]{Branislav Sitar} 
\author[Comenius]{Peter Strmen\fnref{1}} 
\author[GSI]{Maya Takechi}
\author[RCNP]{Junki Tanaka}
\author[Santiago]{Jossitt Vargas} 
\author[GSI]{Helmut Weick}
\author[GSI]{John Stuart Winfield\fnref{1}}

\cortext[cor1]{Corresponding Author}
\address[bh]{School of Physics, Beihang University, Beijing 100191, China}
\address[RCNP]{Research Center for Nuclear Physics (RCNP), Osaka University, Osaka 567-0047, Japan}
\address[SaintMary]{Astronomy and Physics Department, Saint Mary’s University, Halifax B3H 3C3, Canada}
\address[TRIUMF]{TRIUMF, Vancouver V6T 4A3, Canada}
\address[GSI]{GSI Helmholtzzentrum f\"ur Schwerionenforschung, Darmstadt D-64291, Germany}
\address[Justus-Liebig]{Justus-Liebig University, Gießen 35392, Germany}
\address[HFHF]{Helmholtz Research Academy Hesse for FAIR (HFHF), GSI Helmholtz Center for Heavy Ion Research, Gießen 35392, Germany}
\address[Santiago]{Universidad de Santiago de Compostela, Santiago de Compostella E-15706, Spain}
\address[IIT]{Department of Physics, Indian Institute of Technology (Indian School of Mines), Dhanbad 826004, India}
\address[Comenius]{Faculty of Mathematics and Physics, Comenius University, Bratislava 84215, Slovakia}
\address[Dalhousie]{Department of Physics and Atmospheric Science, Dalhousie University, Halifax B3H 4R2, Canada}

\fntext[1]{Deceased.}

\begin{abstract}
We report the charge-changing cross sections ($\sigma_{\text{cc}}$) of 24 $p$-shell nuclides on both hydrogen and carbon at about 900$A$ MeV, of which $^{8,9}$Li, $^{10\textendash12}$Be, $^{10,14,15}$B, $^{14,15,17\textendash22}$N and $^{16}$O on hydrogen and $^{8,9}$Li on carbon are for the first time.
Benefiting from the data set, 
we found a new and robust relationship between the scaling factor of the Glauber model calculations and the separation energies of the nuclei of interest on both targets.
This allows us to deduce proton radii ($R_p$) for the first time from the cross sections on hydrogen.
Nearly identical $R_p$ values are deduced from both target data for the neutron-rich carbon isotopes, however, the $R_p$ from the hydrogen target is systematically smaller in the neutron-rich nitrogen isotopes.
This calls for further experimental and theoretical investigations.
\end{abstract}

\end{frontmatter}

\section{Introduction}
\label{Section:Intro}

Uncovering how protons and neutrons distribute in atomic nuclei is one of the keys to understanding the strong force that binds nucleons together, the nature of new phenomena occurring in exotic nuclei, and the Equation of State (EOS) of nuclear matter that determines the properties of astronomical objects.
Two of the most quoted static properties for nuclear extension in space are 
the root-mean-square ($rms$) point-matter radius ($R_m$) and $rms$ point-proton radii ($R_p$), reflecting the distribution of nucleons and protons in space, respectively. 
Methods using the electroweak probe via, \textit{e.g.}, electron scattering and isotopic shift, and the hadronic probe via, \textit{e.g.}, proton elastic scattering, interaction or charge-changing reactions, have been developed to determine the nuclear size~\cite{Angel2013ADNDT,Tanihata2013PPNP,Sakaguchi2017PPNP,suda_prospects_2017}. 
In the last decades, the advancement of radioactive ion beams allows studying nuclear reactions in inverse kinematics involving short-lived nuclei with good precision. 
This makes the hadronic probe increasingly important to study nuclear sizes up to the most exotic nuclei.

Nucleon collisions at hundreds to thousands of MeV/nucleon represent a major application scenario for hadronic probes. 
Precise measurements of interaction, charge-changing, and neutron-removal cross sections,
especially for the very short-lived and weak-intensity nuclides close to the neutron drip line,
have contributed significantly to our understanding of halo nuclei~\cite{Tanih1985PRL}, neutron skin~\cite{ozawa2001nuclear,bagchi2019neutron,Tanaka2020PRL}, shell structure~\cite{Ozawa00PhysRevLett.84.5493,kanungo2016proton,kaur2022proton}, and the equation of state (EOS) of nuclear matter \cite{aumann2017peeling,xu2022constraining}.
In particular, the proton radii (or charge radii) of a significant fraction of these exotic nuclides cannot be determined by electron scattering and isotope shift due to the inherent limitations of these methods.

In contrast to the electroweak probe, where the electroweak interaction is accurately known, the hadronic probe, however, may suffer from uncertainties due to the complex nature of strong nuclear force, which affects both the structure of collided nuclei and the reaction dynamics.
The Glauber model, which describes the charge-changing reaction by considering the interactions of protons of the projectile nuclide with protons and neutrons of the target nuclide, has been used to deduce the $rms$ proton radius of the projectile nuclide from the charge-changing cross section.
In this approach, one often needs to introduce a sort of normalization to reproduce the cross sections of stable nuclei on the relevant target since the calculations are systematically smaller than the experimental data~\cite{yamaguchi2010energy,yamaguchi2011scaling,bagchi2019neutron, Wang_2023}. 
The reason behind this is not clear yet, but possible mechanisms have been proposed, including the proton evaporation (PE) process after neutron(s) removal \cite{Tanaka2022PRC.106.014617, Zhao2023} and the `$p$\textendash$n$ exchange'~\cite{suzuki2016parameter}.
Moreover, the data on carbon and hydrogen targets may require different considerations~\cite {webber1990total,ozawa2014charge}.
Therefore, determining the fact responsible for the systematic difference between the experimental data and Glauber model interpretation is crucial for computing the proton radius of nuclides, especially exotic ones.

In the present paper, we report the charge-changing cross sections of 24 $p$-shell nuclei from Li to O isotopes at about 900$A$ MeV on carbon ($\sigma_{\text{cc,C}}$) and hydrogen targets ($\sigma_{\text{cc,H}}$) together, 
where $^{8,9}$Li, $^{10\textendash12}$Be, $^{10,14,15}$B, $^{14,15,17\textendash22}$N and $^{16}$O on hydrogen and $^{8,9}$Li on carbon target are reported for the first time.
Other data have been published already in Refs.~\cite{suzuki2016parameter,terashima2014proton,estrade2014proton,kanungo2016proton,bagchi2019neutron,kaur2022proton} but we reanalyzed independently here.
A novel systematic behavior of the scaling factors that include isotope dependence is presented. 
Proton radii determined with the new method at both two targets are presented together with matter radii.

\section{Experiment and Result}
\label{Section:ExpResults}

The experiment was performed with the fragment separator FRS at GSI~\cite{geissel1992gsi}, Germany.
The isotopes of interest were produced by fragmentation of 1$A$ GeV $^{22}$Ne and $^{40}$Ar ions in a 6.3 g/cm$^2$ Be target, then separated and identified in flight on the event-by-event basis by the magnetic rigidity ($B\rho$), time of flight (TOF), and energy loss (${\Delta}E$) measurements. 
The $\sigma_{\text{cc}}$ measurements were carried out with the transmission method on a 4.01 g/cm$^2$ carbon target and a 3.375 g/cm$^2$ polyethylene target. 
The cross sections on hydrogen were derived by subtracting the cross section on the C target from those for the polyethylene target.
The method of extracting cross sections is described in the previous paper \cite{kanungo2016proton}.
One important difference of the cross section treatment has to be noted. 
A veto counter with a central square was placed just in front of the reaction target primarily to remove multi-hit and scattered light particles from upstream.
With the C target data, the cross sections deduced using the veto information are systematically smaller than those without by about 6\%, but this induces a minor effect in the extraction of the published $R_p$ value~\cite{bagchi2019neutron}. 
Moreover, the relevant cross sections on the H target do not change with the veto information. 
This effect occurs because the veto likely detects some light particle events from the back-scattering of the target fragmentation, and simulations validate this impact.
We use here the cross section data without considering the veto.

Newly determined cross sections in the H target are shown in Table~\ref{tab:CCCS} together with some related quantities.
To extract $R_p$ from $\sigma_{\text{cc}}$, we apply the zero-range optical-limit approximation (ZROLA) Glauber model. 
It employs the Eikonal approximation, which is valid in the experimental energy domain and is able to accurately represent the interaction cross sections \cite{Tanihata2013PPNP}.
In this approach, only the collisions of the projectile protons with the target nuclide contribute to $\sigma_{\text{cc}}$, while the projectile neutrons are treated as spectators.
The only inputs to the model are the known nucleon-nucleon ($NN$) cross sections, the proton density distribution of projectile nuclide, and the proton and neutron density of target nuclide. 
We also performed the finite-range (FROLA) model~\cite{horiuchi2007systematic,abu2008reaction} and eventually achieved the identical radius results.
This indicates that our approach is robust and independent of both models.

\begin{table}[htbp]
\centering
\setlength{\tabcolsep}{1.5pt}
\caption{\label{tab:CCCS}Separation energy $S_{\!1}$, secondary beam energies at the middle of the target, measured $\sigma_{\text{cc,C}}$, $\sigma_{\text{cc,H}}$ and the ratios of the measured CCCS to the ZROLA calculation.}
\begin{tabular}{l c c l l l l l}
\hline
Iso- & $S_{\!1}$ & $E/A$ & $\sigma_{\text{cc,C}}^{\text{exp}}$ & $\sigma_{\text{cc,H}}^{\text{exp}}$ & \multicolumn{2}{c}{$\sigma_{\text{cc}}^{\text{exp}}/\sigma_{\text{cc}}^{\text{calc}}$} \\
\cline{6-7}
tope&(MeV)&(MeV)&(mb)&(mb)& Carbon & Hydrogen \\\hline

$^{8}$Li&12.0&901&642(16)&149(14)&1.184(29)&1.418(133)\\
$^{9}$Li&16.5&958&608(14)&141(8)&1.151(27)&1.434(88)\\
$^{10}$Be&23.7&994&713(11)&206(10)&1.185(18)&1.596(77)\\
$^{11}$Be&20.1&928&713(15)&179(13)&1.147(24)&1.343(98)\\
$^{12}$Be&23.3&959&741(11)&186(7)&&\\
$^{10}$B&8.2&930&784(10)&233(8)&1.190(15)&1.526(52)\\
$^{14}$B&16.8&990&751(14)&223(9)&&\\
$^{15}$B&20.0&963&773(3)&206(2)&&\\
$^{12}$C&27.4&928&764(5)&217(4)&1.080(7)&1.246(23)\\
$^{14}$C&25.7&991&797(9)&220(8)&1.111(13)&1.246(45)\\
$^{15}$C&22.0&893&791(3)&233(3)&&\\
$^{16}$C&25.2&808&809(17)&231(16)&&\\
$^{17}$C&23.2&962&800(9)&232(7)&&\\
$^{18}$C&27.5&955&813(5)&226(4)&&\\
$^{19}$C&26.6&880&807(8)&224(6)&&\\
$^{14}$N&12.5&924&868(6)&280(5)&1.142(8)&1.421(25)\\
$^{15}$N&18.4&762&861(19)&277(30)&1.117(25)&1.395(151)\\
$^{17}$N&17.4&927&874(6)&269(6)&&\\
$^{18}$N&15.9&848&869(14)&265(19)&&\\
$^{19}$N&20.6&949&860(4)&256(4)&&\\
$^{20}$N&18.5&877&866(3)&257(3)&&\\
$^{21}$N&22.5&874&846(9)&245(9)&&\\
$^{22}$N&20.9&882&858(15)&244(12)&&\\
$^{16}$O&23.0&920&898(7)&284(7)&1.088(8)&1.269(31)\\

\hline
\end{tabular}
\end{table}

\section{New approach for deducing $rms$ proton radii}
\label{DISCUSSIONS}
We computed $\sigma_{\text{cc}}^{\text{calc}}$ for ten nuclei ($^{8,9}$Li, $^{10,11}$Be, $^{10}$B, $^{12,14}$C, $^{14,15}$N, and $^{16}$O) at 900$A$ MeV on C and H targets with ZROLA.
The ten nuclei have well-determined charge radii $R_{\text{c}}$~\cite{Angel2013ADNDT},
from which one can deduce $R_p$ and the harmonic oscillator (HO) density distributions.
The density in the Dirac delta form is utilized for the hydrogen target, while the HO density distribution is for the carbon target.

The ratios of the measured charge-changing cross sections to the ZROLA calculations are shown in Table~\ref{tab:CCCS}. 
The ZROLA values at 900$A$ MeV are systematically lower than our data on carbon by 10\%\textendash20\% and than the hydrogen data by 20\%\textendash50\%. 
Such underestimation in cross sections, which was also observed at 300$A$ MeV~\cite{yamaguchi2011scaling,Tanaka2022PRC.106.014617,Zhao2023,ozawa2014charge, Wang_2023}, is considered to be due to the fact that the Glauber model only reflects the direct proton removal process of the projectile, $\sigma_{\text{cc,Direct}}$.
In reality, there is a possibility that charged particle(s), mostly protons, can be emitted following the direct neutron removals, where the pre-fragment is highly excited. This process can contribute to the final $\sigma_{\text{cc}}^{\text{exp}}$~\cite{Tanaka2022PRC.106.014617,Zhao2023}, but is beyond the current reaction model. 

\begin{figure}[htbp]
\centering
\includegraphics[width=0.45\textwidth]{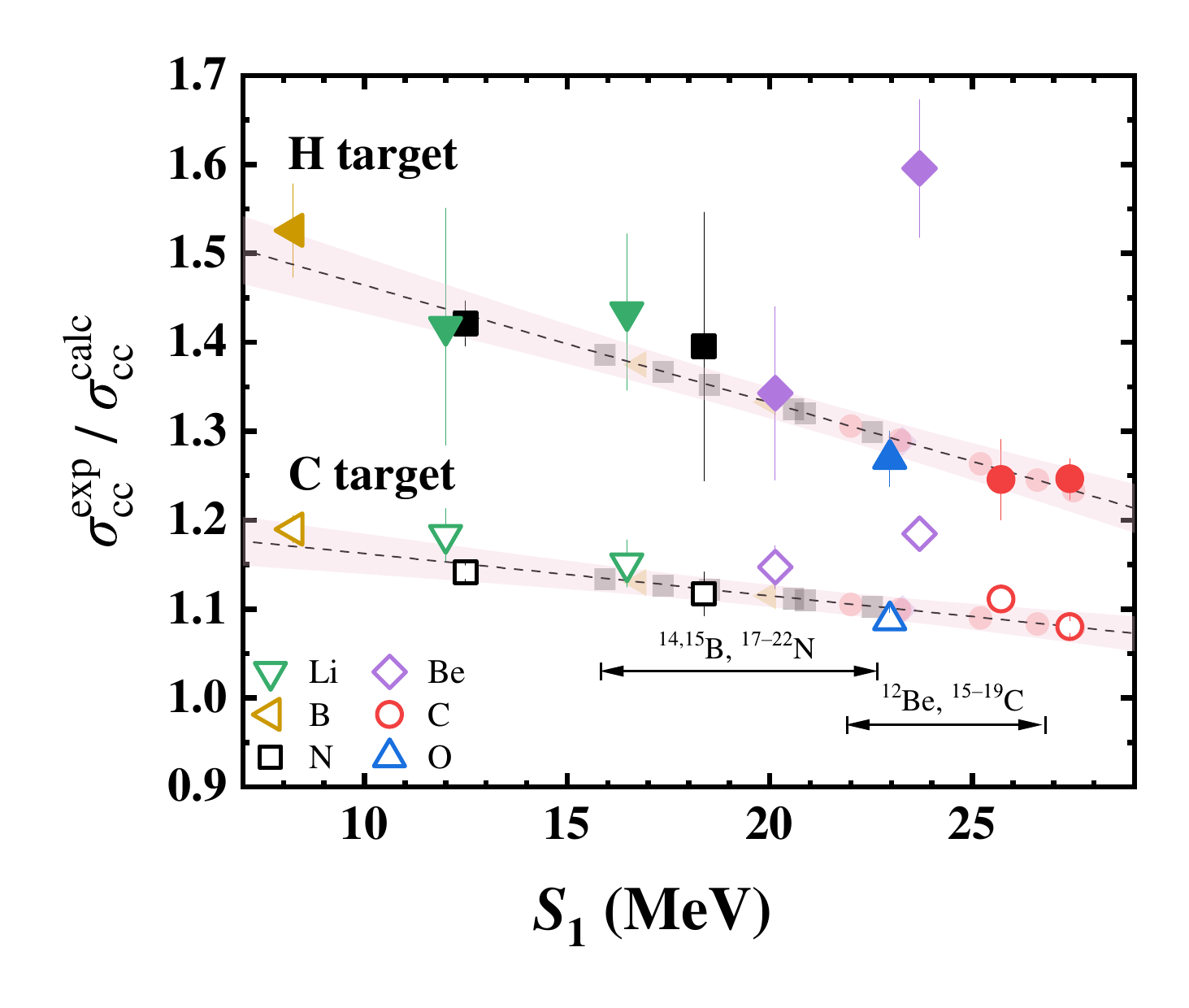}
\caption{\label{fig:ratio} 
Ratios of experimental charge-changing cross sections to theoretical ones,
$\sigma_{\text{cc}}^{\text{exp}}/\sigma_{\text{cc}}^{\text{calc}}$, as a function of the separation energy, $S_{\!1}$ for ten nuclei with well known $R_p$. The data on the H target and C target are indicated with filled and open marks, respectively. 
The dashed line and shaded band represent the best linear fit to the data but excluding $^{10}$Be, and the 95\% confidence interval.
Indicated along the lines are also the positions of the remaining 14 neutron-rich isotopes by translucent symbols.}
\end{figure}

To better account for the PE process following the projectile neutron removal, 
we introduce a new quantity, $S_{\!1}(^A_NZ) \equiv S_{\!n}(^A_NZ)+S_{\!p}(^{A-1}_{N-1}Z)$.
$S_{\!n}(^A_NZ)$ is the neutron separation energy from the projectile nucleus, and the $S_{\!p}(^{A-1}_{N-1}Z)$ is the proton separation energy from the one-neutron removed nucleus.
We sort the $\sigma_{\text{cc}}^{\text{exp}}/\sigma_{\text{cc}}^{\text{calc}}$ values of stable nuclei according to $S_{\!1}(^A_NZ)$ in Fig.~\ref{fig:ratio}. 
It is interesting to see that the ratios on each target show a linear dependence on $S_{\!1}(^A_NZ)$, except $^{10}$Be. 
The linearly fitted lines for the hydrogen and carbon data are $f(\mathrm{H})=-0.0132(10) S_{\!1} + 1.596(23)$ and $f(\mathrm{C})=-0.0047(8) S_{\!1} + 1.209(16)$, respectively.
The H target data are more sensitive on $S_{\!1}(^A_NZ)$ than the relevant C target data.
Such correlation cannot be seen when plotting $\sigma_{\text{cc}}^{\text{exp}}/\sigma_{\text{cc}}^{\text{calc}}$ versus $Z/N$ \cite{yamaguchi2011scaling} or other parameters.
Deviation of $^{10}$Be from this line is easily understood because the neutron-removed nucleus $^{8}$Be decays immediately into a pair of $\alpha$ particles after direct or cascade two neutron removal \cite{terashima2014proton,warner2001total,hue2017neutron}.
This will result in a sizable charge change but is not included in $S_{\!1}$. 
It should be noted, however, that there are rare nuclei with the $\alpha$-evaporation probability after neutron removals, like $^{10}$Be. 
In the dominant cases, the proton-evaporation process occurs after neutron removals in high-energy heavy-ion collisions.

The distinct correlation identified in Fig.~\ref{fig:ratio} is the first evidence of the widespread presence of the PE process in charge-changing reactions. 
This demonstrates that it is a universal process, occurring not only in (near-)stable isotopes but also in neutron-rich isotopes.
Theoretical calculations of $(p, pn)$ reactions suggest that decreasing separation energies $S_{\!n}$ lead to increasing neutron removal cross sections~\cite{aumann2013quasifree,aumann2021quenching,cravo2016distortion}.
$S_{\!n}$, the first term of $S_{\!1}$, thus is expected to be positively correlated with the one-neutron removal process, the leading part in the total neutron-removals.
In addition, a larger $S_{\!n}(^{A}_{N}Z)$ may lead to a lower excitation energy for the pre-fragment $^{A-1}_{N-1}Z$ produced right after the one-neutron removal.
As a result, it becomes less probable for the hot pre-fragment to de-excite via charged particles.
When extended to the collision systems with different projectiles and beam energies, the orbital of the removed neutron and the beam energy may also affect the excitation of pre-fragments.
Protons are most favorable among charged particles to be emitted due to the Coulomb barrier, and their emission probability depends on the excitation energy and the proton emission threshold $S_{\!p}(^{A-1}_{N-1}Z)$, the second term of $S_{\!1}$. 
The observed trend in Fig.~\ref{fig:ratio} shows consistent behavior with the PE process. 
One can further extend $S_{\!1}$ by including multiple-nucleon removal and evaporation processes, but we do not see significant improvement in the fit. 
This is mainly due to the cross sections for removing two or more neutrons being much smaller than those for one neutron \cite{kobayashi2012one} and the experimental uncertainties.

The well-defined correlation of $\sigma_{\text{cc}}^{\text{exp}}/\sigma_{\text{cc}}^{\text{calc}}$ with $S_{\!1}$ offers an empirical way to calibrate the model and deduce $R_p$ for the 14 nuclides ($^{12}$Be, $^{14,15}$B, $^{15\textendash19}$C, $^{17\textendash22}$N), whose radii were not determined by the electroweak interaction probe. 
The important thing is that the systematic obtained by the $R_p$ known nuclei spread in a wide range of $S_{\!1}$ so that the scaling value for the exotic nuclei can be obtained by the interpolation of the systematic, instead of the extrapolation. 
We incorporate this phenomenological scaling factor $f(S_{\!1},\text{Targ})$ into the ZR Glauber model, where $f$ is determined by the $S_{\!1}$ of the projectile and the type of target. 
This scaling factor introduces uncertainty by typically 1.3\% and 2.4\% for $\sigma_{\text{cc}}^{\text{calc}}$ on C and H targets, respectively.
The HO type proton density distribution and proton radius $R_p$ of the projectile nucleus are determined when matching $\sigma_{\text{cc}}^{\text{calc}}$ to $\sigma_{\text{cc}}^{\text{exp}}$.
Once $R_p$ is determined, we can calculate the neutron radius $R_n$ as well as the matter radius $R_m$ in a similar manner but by reproducing instead the interaction cross sections at similar beam energies ($\sigma_{\text{I}}$)~\cite{ozawa2001nuclear}.

\begin{figure}[htb]
\centering
\includegraphics[width=0.45\textwidth]{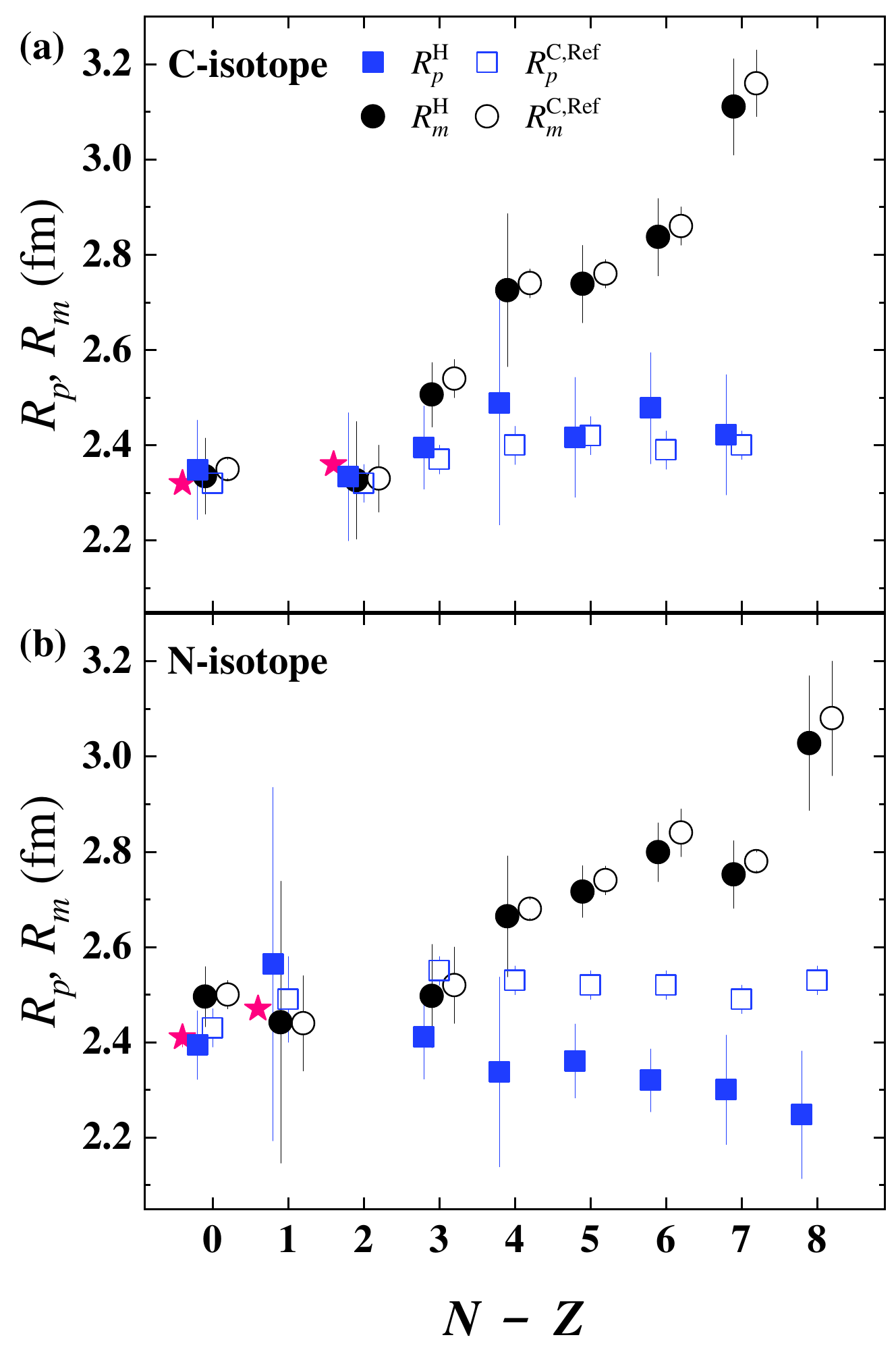}
\caption{\label{fig:rprmall} 
    $R_p$ extracted from $\sigma_{\text{cc,H}}$ ($R_p^{\text{H}}$, filled squares) and $R_m$ from $\sigma_{\text{I}}$ \cite{ozawa2001nuclear} ($R_m^{\text{H}}$, filled circles), compared with the published radii \cite{kanungo2016proton,bagchi2019neutron} on the C target ($R_p^{\text{C,Ref}}$, open squares, $R_m^{\text{C,Ref}}$, open circles) for C (a) and N isotopes (b). 
    The tabulated $R_p$ values for $^{12,14}$C and $^{14,15}$N~\cite{Angel2013ADNDT} are shown by stars. The $N-Z$ values for the data have been slightly shifted for better visibility. 
}
\end{figure}

We extracted the proton and matter radius of all nuclides from the C target ($R_p^{\text{C}}$, $R_m^{\text{C}}$) and the H target ($R_p^{\text{H}}$, $R_m^{\text{H}}$), where radii are extracted for the first time from H target.
For $^{12,14\textendash19}$C and $^{14,15,17\textendash22}$N isotopes, the $R_p^{\text{C}}$ and $R_m^{\text{C}}$ agree well with the results of previous publications ($R_p^{\text{C,Ref}}$, $R_m^{\text{C,Ref}}$), within errors \cite{kanungo2016proton,bagchi2019neutron}.
In Fig.~\ref{fig:rprmall}, we compare the radii on the H target with those on the C target that have been published.

Fig.~\ref{fig:rprmall}a displays the $rms$ matter and the proton radii of carbon isotopes determined by the C target and H target. 
Excellent agreement is seen for both $R_p$ and $R_m$.
Fig.~\ref{fig:rprmall}b shows the radii of nitrogen isotopes. 
The $R_m$ determined from both again agree very well.
However, the $R_p$ values determined by the C target tend to be systematically larger than the according H target ones when moving to the neutron-rich side.

We would like to emphasize the new scaling identified in Fig.~\ref{fig:ratio} is also presented when using the FROLA models instead. 
In fact, the FROLA calculations are systematically larger than the ZROLA by about 3\% and 7\% for the C and H targets, respectively. 
However, this systematic difference can be incorporated into the scaling factor function when compared to the same experimental data.
In the FROLA case, the relevant fits for the C and H target data are $f(\mathrm{C}) = -0.0127(9) S_{\!1} + 1.542(18)$ and $f(\mathrm{H}) = -0.0045(7) S_{\!1} + 1.169(16)$, respectively.
Employing FROLA will give almost identical slopes but lower intercepts than employing ZROLA. Eventually, one obtains consistent $R_p$ values which are independent of the Glauber models.

\begin{figure}[htbp]
\centering
\includegraphics[width=0.45\textwidth]{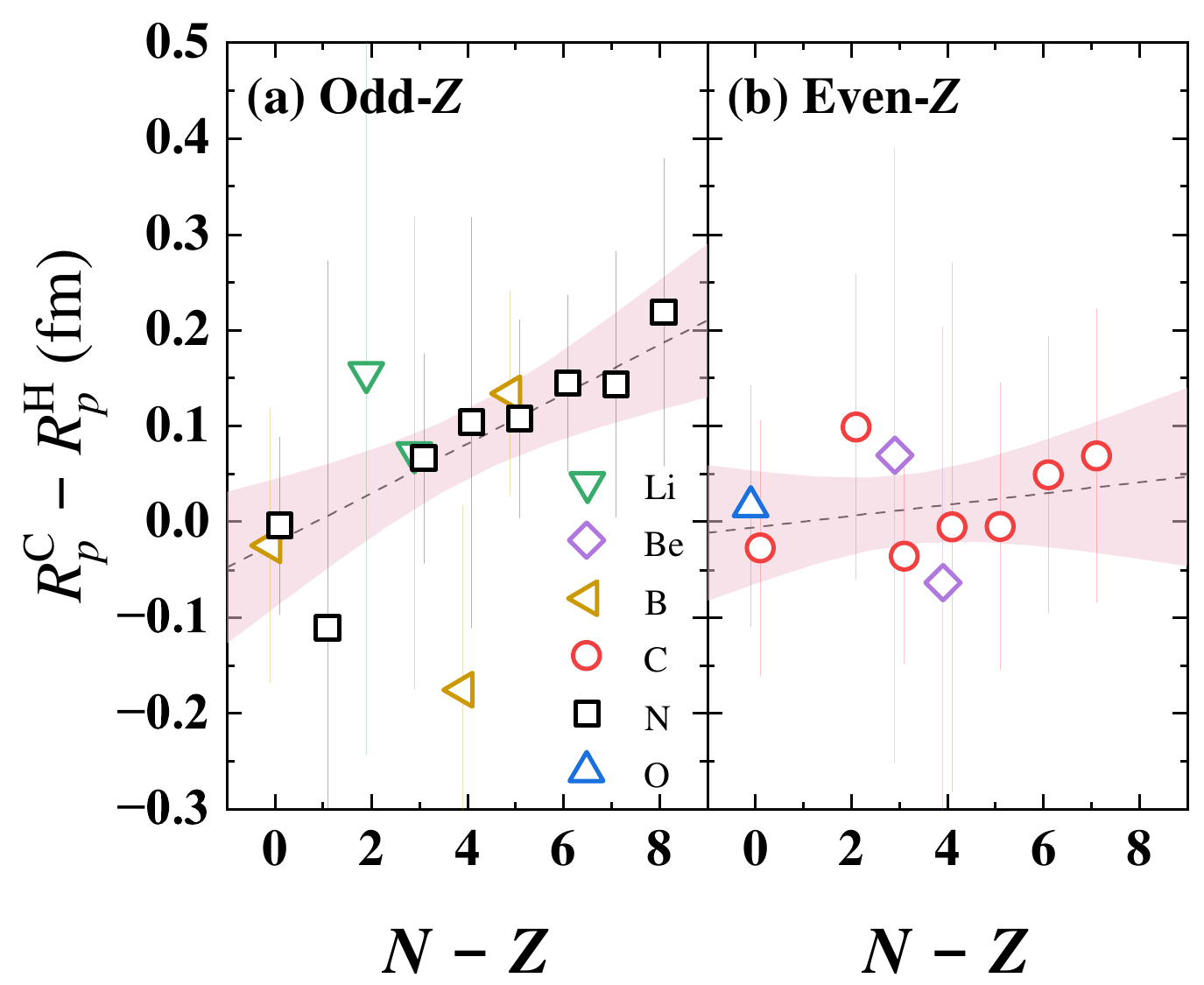}
\caption{
\label{fig:oddeven1} Difference of the proton radii extracted from the carbon and hydrogen data, $R_p^{\text{C}} - R_p^{\text{H}}$, as a function of $N - Z$ for odd-$Z$ (a) and even-Z (b) isotopes.
The dashed line shows the best linear fit to the data, while the shaded band indicates the 95\% confidence interval. The $N-Z$ values for the data are slightly shifted. 
Note the combined standard uncertainties of $R_p^{\text{C}}-R_p^{\text{H}}$ are shown for each point.
}
\end{figure}

A conclusive explanation for the observed difference in the deduced $R_p$ for neutron-rich N isotopes remains open.
Knockout reactions have indicated that a proton as a target can observe both the inside and surface of the projectile, whereas a carbon as a target is more sensitive to the surface \cite{aumann2013quasifree,aumann2021quenching}.
In this sense, the shape of the projectile's proton density distribution can play a role since the tail of the proton density distribution displays greater sensitivity to the C target than to the H target.
It thus could enhance the effect for the N isotopes having an odd number of protons. 
In Fig.~\ref{fig:oddeven1}, we summarize the $R_p^{\text{C}} - R_p^{\text{H}}$ for all the isotopes in this work as a function of neutron excess. Note that the nuclei with $N-Z=0,1,2$ and part of 3 have been used to calibrate our approach to deduce proton radii. Overall, the differences follow approximately a linear trend with neutron excess. 
For odd-$Z$ data, the best-fit straight line has a slope of 0.025(6), which is significantly larger than 0.006(6) for the even-$Z$ data.
This may indicate that employing the HO type to characterize the proton density distribution of neutron-rich isotopes of odd-$Z$ nuclides is insufficient.

\section{Summary}
\label{sum}

In summary, we report here the analysis of data on the $\sigma_{\text{cc}}$ of 24 $p$-shell nuclides on carbon and hydrogen at around 900$A$ MeV. 
We found a new correlation of $\sigma_{\text{cc}}$ with $S_{\!1}$, an index for the proton-evaporation process after neutron removals in charge-changing reactions. With this correlation, we can determine the isotope-dependent scaling factor empirically. 
This 
provides a new insight 
to determine $R_p$ from $\sigma_{\text{cc}}$ on either a proton or any heavy-ion targets to study nuclei far from the stability line in experiments using inverse kinematics.

We are aware that the correlation with $S_{\!1}$ is empirical. 
It remains a challenge for the theory to quantify the link between the PE process with $S_{\!1}$.
Other mechanisms, such as the excitation of giant resonances via the nuclear interaction~\cite{bertulani2019neutron,teixeira2022nuclear} and the `$p$\textendash$n$ exchange' process~\cite{suzuki2016parameter}, may also partially affect the charge-changing reaction.
Dedicated theoretical calculations of light nuclei densities and momentum distributions~\cite{Piaru2023PhysRevC.107, sun2018shrunk, sun2021rotating} will be very interesting.
This will help to pin down the factors responsible for the difference observed by the two probes and to understand how protons distribute inside the neutron skin. 
Furthermore, extending the $S_{\!1}$ scaling method beyond the $p$-shell nuclei will need further evaluation by a large data set composed of different projectiles, incident energies, and targets. 
In the forthcoming experiments \cite{SUN201878}, we will attempt to identify the PE process and to what extent its contribution to the $\sigma_{\text{cc}}$.
Nevertheless, this empirical correlation hints at a deeper understanding. 
The present analysis shows that there are excellent agreements in the deduced $R_p$ values of C isotopes or even-$Z$ isotopes from C-target and H-target data, but there seem to be systematic discrepancies in the case of N isotopes or odd-$Z$ isotopes. 
This may related to the basic concept of how to characterize the nuclear size and the possible effect of the heavy-ion probes.

\section*{Conflict of interest}
\label{Conflictofinterest}
The authors declare that they have no conflict of interest.

\section*{Author contributions}
\label{Authorcontributions}

Bao-Hua Sun and Isao Tanihata supervised the project. 
JiChao Zhang, Feng Wang, Satoru Terashima, Bao-Hua Sun, and Isao Tanihata processed the data. 
JiChao Zhang and Bao-Hua Sun performed calculations. 
JiChao Zhang, BaoHua Sun, and Isao Tanihata wrote the manuscript with input from all authors, and Rituparna Kanungo and Christoph Scheidenberger were deeply involved in the discussion and writing process.
Rituparna Kanungo and Isao Tanihata proposed and led the experiment.
All authors participated in the experiments, data processing, and manuscript revision.

\section*{Acknowledgments}
\label{acknowledgments}
The authors are thankful for the support of the GSI accelerator staff and the FRS technical staff for the efficient preparation of the experiment setup.
The current analysis was performed in FAIR Phase\textendash0. This work is partly supported by the National Natural Science Foundation of China under Contracts No. 12325506, and No. 11961141004, and the ``111 center" under Grant No. B20065. 
The support from NSERC, Canada, for this work is gratefully acknowledged. 
The support of the Faculty Research Scheme at IIT (ISM) Dhanbad (Grant No. FRS(154)/2021-2022/Physics) is gratefully acknowledged.
The support of the PR China government and Beihang University under the Thousand Talent program is gratefully acknowledged. 
We thank Prof. W. Horiuchi, Prof. Carlos Bertulani, Dr. X Roca Maza, and Dr. Xiang-xiang Sun for the helpful discussions.


\begin{thebibliography}{10}

\bibitem{Angel2013ADNDT}
Angeli I, Marinova KP.
\newblock Table of experimental nuclear ground state charge radii: An update.
\newblock At Data Nucl Data Tables 2013;99:69--95.

\bibitem{Tanihata2013PPNP}
Tanihata I, Savajols H, and Kanungo R.
\newblock Recent experimental progress in nuclear halo structure studies.
\newblock {Prog Part Nucl Phys} 2013;68:215--313.

\bibitem{Sakaguchi2017PPNP}
Sakaguchi H, Zenihiro J.
\newblock Proton elastic scattering from stable and unstable
  nuclei-extraction of nuclear densities.
\newblock {Prog Part Nucl Phys} 2017;97:1--52.

\bibitem{suda_prospects_2017}
Suda T, Simon H.
\newblock Prospects for electron scattering on unstable, exotic nuclei.
\newblock {Prog Part Nucl Phys} 2017;96:1--31.

\bibitem{Tanih1985PRL}
Tanihata I, Hamagaki H, Hashimoto O, et~al.
\newblock Measurements of interaction cross sections and nuclear radii in the
  light \textit{p}-shell region.
\newblock Phys Rev Lett 1985;55:2676.

\bibitem{ozawa2001nuclear}
Ozawa A, Suzuki T, Tanihata I.
\newblock Nuclear size and related topics.
\newblock Nucl Phys A 2001;693:32--62.

\bibitem{bagchi2019neutron}
Bagchi S, Kanungo R, Horiuchi W, et~al.
\newblock Neutron skin and signature of the $N = 14$ shell gap found from
  measured proton radii of $^{17\textendash22}$N.
\newblock Phys Lett B 2019;790:251--256.

\bibitem{Tanaka2020PRL}
Tanaka M, Takechi M, Homma A, et~al.
\newblock Swelling of doubly magic $^{48}$Ca core in ca isotopes beyond $N = 28$.
\newblock Phys Rev Lett 2020;124:102501.

\bibitem{Ozawa00PhysRevLett.84.5493}
Ozawa A, Kobayashi T, Suzuki T, et~al.
\newblock New magic number, $N = 16$, near the neutron drip line.
\newblock Phys Rev Lett 2000;84:5493.

\bibitem{kanungo2016proton}
Kanungo R, Horiuchi W, Hagen G, et~al.
\newblock Proton distribution radii of $^{12\textendash19}$C illuminate features of neutron
  halos.
\newblock Phys Rev Lett 2016;117:102501.

\bibitem{kaur2022proton}
Kaur S, Kanungo R, Horiuchi W, et~al.
\newblock Proton distribution radii of $^{16\textendash24}$O: Signatures of new shell
  closures and neutron skin.
\newblock Phys Rev Lett 2022;129:142502.

\bibitem{aumann2017peeling}
Aumann T, Bertulani CA, Schindler F, et~al.
\newblock Peeling off neutron skins from neutron-rich nuclei: Constraints on
  the symmetry energy from neutron-removal cross sections.
\newblock Phys Rev Lett 2017;119:262501.

\bibitem{xu2022constraining}
Xu JY, Li ZZ, Sun BH, et~al.
\newblock Constraining equation of state of nuclear matter by charge-changing
  cross section measurements of mirror nuclei.
\newblock Phys Lett B 2011;833:137333.

\bibitem{yamaguchi2010energy}
Yamaguchi T, Fukuda M, Fukuda S, et~al.
\newblock Energy-dependent charge-changing cross sections and proton
  distribution of $^{28}$Si.
\newblock Phys Rev C 2010;82:014609.

\bibitem{yamaguchi2011scaling}
Yamaguchi T, Hachiuma I, Kitagawa A, et~al.
\newblock Scaling of charge-changing interaction cross sections and
  point-proton radii of neutron-rich carbon isotopes.
\newblock Phys Rev Lett 2011;107:032502.

\bibitem{Wang_2023}
Wang CJ, Guo G, Ong HJ, et~al.
\newblock Charge-changing cross section measurements of 300 MeV/nucleon $^{28}$Si on
  carbon and data analysis.
\newblock Chin Phys C 2023;47:084001.

\bibitem{Tanaka2022PRC.106.014617}
Tanaka M, Takechi M, Homma A, et~al.
\newblock Charge-changing cross sections for $^{42\textendash51}$Ca and effect of
  charged-particle evaporation induced by neutron-removal reactions.
\newblock Phys Rev C 2022;106:014617.

\bibitem{Zhao2023}
Zhao JW, Sun BH, Tanihata I, et~al.
\newblock Isospin-dependence of the charge-changing cross-section shaped by the
  charged-particle evaporation process.
\newblock Phys Lett B 2023;847:138269.

\bibitem{suzuki2016parameter}
Suzuki Y, Horiuchi W, Terashima S, et~al.
\newblock Parameter-free calculation of charge-changing cross sections at high
  energy.
\newblock Phys Rev C 2016;94:011602.

\bibitem{webber1990total}
Webber WR, Kish JC, and Schrier DA.
\newblock Total charge and mass changing cross sections of relativistic nuclei
  in hydrogen, helium, and carbon targets.
\newblock Phys Rev C 1990;41:520.

\bibitem{ozawa2014charge}
Ozawa A, Moriguchi T, Ohtsubo T, et~al.
\newblock Charge-changing cross sections of $^{30}$Ne, $^{32,33}$Na with a proton
  target.
\newblock Phys Rev C 2014;89:044602.

\bibitem{terashima2014proton}
Terashima S, Tanihata I, Kanungo R, et~al.
\newblock Proton radius of $^{14}$Be from measurement of charge-changing cross
  sections.
\newblock Prog Theor Exp Phys 2014;2014:101D02.

\bibitem{estrade2014proton}
Estrad{\'e} A, Kanungo R, Horiuchi W, et~al.
\newblock Proton radii of $^{12\textendash17}$B define a thick neutron surface in $^{17}$B.
\newblock Phys Rev Lett 2014;113:132501.

\bibitem{geissel1992gsi}
Geissel H, Armbruster P, Behr KH, et~al.
\newblock The GSI projectile fragment separator (FRS): a versatile magnetic
  system for relativistic heavy ions.
\newblock Nucl Instrum Methods Phys Res B 1992;70:286--297.

\bibitem{horiuchi2007systematic}
Horiuchi W, Suzuki Y, Ibrahim BA, et~al.
\newblock Systematic analysis of reaction cross sections of carbon isotopes.
\newblock Phys Rev C 2007;75:044607.

\bibitem{abu2008reaction}
Ibrahim BA, Horiuchi W, Kohama A, et~al.
\newblock Reaction cross sections of carbon isotopes incident on a proton.
\newblock Phys Rev C 2008;77:034607.

\bibitem{warner2001total}
Warner RE, McKinnon MH, Needleman JS, et~al.
\newblock Total reaction and neutron-removal cross sections of (30--60)\textit{A} MeV
  Be isotopes on Si and Pb.
\newblock Phys Rev C 2001;64:044611.

\bibitem{hue2017neutron}
Hue BM, Isataev T, Lukyanov SM, et~al.
\newblock Neutron-removal cross sections of $^{6,8}$He, $^{8}$Li and $^{9,10}$Be nuclei.
\newblock Eurasian J Phys Funct Mater 2017;1:65--73.

\bibitem{aumann2013quasifree}
Aumann T, Bertulani CA, Ryckebusch J.
\newblock Quasifree $(p,2p)$ and \textit{(p,pn)} reactions with unstable nuclei.
\newblock Phys Rev C 2013;88:064610.

\bibitem{aumann2021quenching}
Aumann T, Barbieri C, Bazin D, et~al.
\newblock Quenching of single-particle strength from direct reactions with
  stable and rare-isotope beams.
\newblock Prog Part Nucl Phys 2021;118:103847.

\bibitem{cravo2016distortion}
Cravo E, Crespo R, Deltuva A.
\newblock Distortion effects on the neutron knockout from exotic nuclei in the
  collision with a proton target.
\newblock Phys Rev C 2016;93:054612.

\bibitem{kobayashi2012one}
Kobayashi N, Nakamura T, Tostevin JA, et~al.
\newblock One- and two-neutron removal reactions from the most neutron-rich
  carbon isotopes.
\newblock Phys Rev C 2012;86:054604.

\bibitem{bertulani2019neutron}
Bertulani CA, Valencia J.
\newblock Neutron skins as laboratory constraints on properties of neutron
  stars and on what we can learn from heavy ion fragmentation reactions.
\newblock Phys Rev C 2019;100:015802.

\bibitem{teixeira2022nuclear}
Teixeira EA, Aumann T, Bertulani CA, et~al.
\newblock Nuclear fragmentation reactions as a probe of neutron skins in
  nuclei.
\newblock Eur Phys J 2022;58:205.

\bibitem{Piaru2023PhysRevC.107}
Piarulli M, Pastore S, Wiringa EB, et~al.
\newblock Densities and momentum distributions in $A \leq 12$ nuclei from chiral
  effective field theory interactions.
\newblock Phys Rev C 2023;107:014314.

\bibitem{sun2018shrunk}
Sun XX, Zhao J, Zhou SG.
\newblock Shrunk halo and quenched shell gap at $N = 16$ in $^{22}$C: Inversion of \textit{sd}
  states and deformation effects.
\newblock Phys Lett B 2018;785:530--535.

\bibitem{sun2021rotating}
Sun XX and Zhou SG.
\newblock Rotating deformed halo nuclei and shape decoupling effects.
\newblock Sci Bull 2021;66:2072--2078.

\bibitem{SUN201878}
Sun BH, Zhao JW, Zhang XH, et~al.
\newblock Towards the full realization of the RIBLL2 beam line at the HIRFL-CSR
  complex.
\newblock Sci Bull 2018;63:78--80.

\end{thebibliography}
\end{document}